\documentclass[%
 aip,
 jmp,%
 amsmath,amssymb,
preprint,%
]{revtex4-1}

\usepackage{graphicx}
\usepackage{dcolumn}
\usepackage{bm}
\usepackage[pdftex,colorlinks=true,bookmarks=false,citecolor=blue,urlcolor=blue]{hyperref}

\begin{document}

\title{High-frequency and high-quality silicon carbide optomechanical microresonators}

\author{Xiyuan Lu}
\address{Department of Physics and Astronomy, University of Rochester, Rochester, NY 14627, USA}
\author{Jonathan Y. Lee}
\address{Department of Electrical and Computer Engineering, University of Rochester, Rochester, NY 14627, USA}
\author{Qiang Lin}
\address{Department of Electrical and Computer Engineering, University of Rochester, Rochester, NY 14627, USA}
\address{Institute of Optics, University of Rochester, Rochester, NY 14627, USA}
\email{qiang.lin@rochester.edu}


\begin{abstract}
Silicon carbide (SiC) exhibits excellent material properties attractive for broad applications. We demonstrate the first SiC optomechanical microresonators that integrate high mechanical frequency, high mechanical quality, and high optical quality into a single device. The radial-breathing mechanical mode has a mechanical frequency up to 1.69 GHz with a mechanical Q around 5500 in atmosphere, which corresponds to a ${\rm f_m\cdot Q_m}$ product as high as $\rm 9.47 \times 10^{12} Hz$. The strong optomechanical coupling allows us to efficiently excite and probe the coherent mechanical oscillation by optical waves. The demonstrated devices, in combination with the superior thermal property, chemical inertness, and defect characteristics of SiC, show great potential for applications in metrology, sensing, and quantum photonics, particularly in harsh environments that are challenging for other device platforms.
\end{abstract}

\maketitle

\section*{Introduction}

Optomechanical resonators couple optical cavities and mechanical resonators mutually through optomechanical interactions mediated by the radiation-pressure forces. With the exceptional capability of probing and controlling mesoscopic mechanical motion down to single quantum level, micro/nano-optomechanical resonators have been intensively investigated in recent years, showing great promise for broad applications in sensing, information processing, time/frequency metrology, and quantum physics \cite{Kippenberg08, Marquart09, Karrai09, Roels10, Poot12, Aspelmeyer14}. To date, diverse optomechanical structures \cite{Aspelmeyer14} have been developed on a variety of material platforms including silica \cite{Vahala05}, silicon nitride \cite{Harris08}, silicon \cite{Painter09}, gallium arsenide \cite{Favero10}, aluminium nitride \cite{Tang12}, diamond \cite{Pernice13}, phospho-silicate glass \cite{Wu13}, and gallium phosphide \cite{Barclay14}. In general, cavity optomechanics relies critically on the underlying device material, requiring not only high optical transparency and large refractive index to support the high-quality and strong-confined optical modes, but also large acoustic velocity and low material damping to support the high-frequency and high-quality mechanical resonances.

Silicon carbide (SiC) is well known for its outstanding thermal, optical, mechanical and chemical properties \cite{Harris95}, with broad applications in high-power electronics, micromechanical sensors, biomedical devices, and astronomical telescopes \cite{Cimalla07, Maboudian13, Robichaud12}. In the past few years, significant efforts have been devoted to develop SiC-based micro/nanophotonic devices \cite{Noda11, Noda12, Lu13, Lipson13, Jelena13, Noda14, Awschalom14, Hu14, Lu14, Lin14, Hu15, Lee15}, greatly attracted by its nonlinear optical properties \cite{Noda14, Lin14} and defect characteristics \cite{Koehl11, Awschalom13}. On the other hand, recent theoretical studies \cite{Kenny08, Ayazi11, Kenny13} show that SiC exhibits intrinsic mechanical quality significantly superior than other materials, with a theoretical frequency-quality (${\rm f_m \cdot Q_m}$) product $\sim 3\times 10^{14}$ at room temperature, due to its exceptionally low phonon-phonon scattering that dominates the intrinsic mechanical loss in the microwave frequency regime. The high intrinsic mechanical quality, together with the outstanding optical properties, makes SiC an excellent material platform for optomechanical applications. Unfortunately, the superior mechanical rigidity and chemical inertness of SiC impose significant challenge on fabricating micro-/nano-photonic devices with high optical and mechanical qualities, which seriously hinders the realization of optomechanical functionalities on the SiC platform.

In this letter, we demonstrate the first SiC optomechanical microresonators that exhibit significant optomechanical coupling with a coefficient up to ${\rm |g_{om}|/2\pi \approx (61 \pm 8)~{\rm GHz/nm}}$, which enables us to efficiently actuate and characterize the mesoscopic mechanical motions by optical means. By optimizing the device structure and the fabrication process, we are able to achieve high optical quality, large mechanical frequency, and high mechanical quality simultaneously in a single device. The whispering-gallery modes exhibit high optical qualities around $\sim 3.8\times10^4$. The radial-breathing mechanical modes show frequencies up to 1.69~GHz and mechanical qualities around 5500. The corresponding ${\rm f_m \cdot Q_m}$ product is $\rm 9.47 \times 10^{12}$, which is the highest value for the fundamental bulk acoustic mode in SiC demonstrated to date \cite{Yang01, Huang03, Ekinci04, Wiser05, Forster06, Jiang06, Azevedo07, Li07, Perisanu07, Yang14, Zhao14, Iacopi14}, to the best of our knowledge.

The high performance of the demonstrated optomechanical microresonators shows that SiC devices are now ready for broad optomechanical applications. With the superior thermal and chemical properties of SiC material\cite{Harris95}, SiC optomechanical devices are particularly attractive for optomechanical sensing, such as displacement, force, mass, and inertial sensing, especially in harsh environments that are challenging for other device platforms. On the other hand, the SiC optomechanical microresonators, in combination with SiC's significant optical nonlinearities \cite{Noda14, Lin14} and unique defect characteristics \cite{Koehl11, Awschalom13}, are of great promise for realizing hybrid micro/nanophotonic circuits for nano-optomechanics, integrated nonlinear photonics, and quantum photonics.

\section*{Results}
\subsection*{Optomechanical device}
\begin{figure}[h]
\centering
\setlength{\belowcaptionskip}{-7pt}
\includegraphics[width=1\columnwidth]{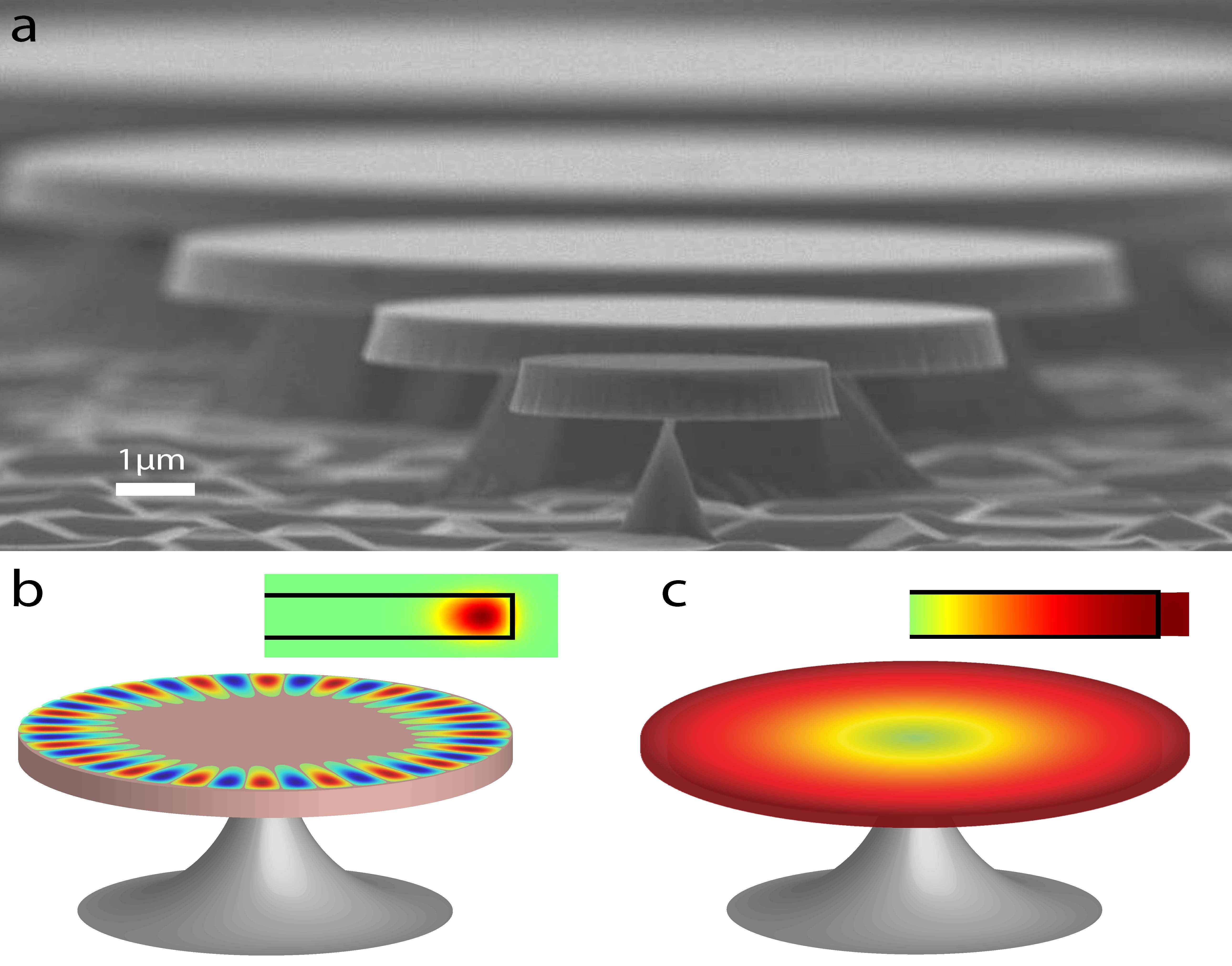}
\caption{(a) Scanning electron microscope (SEM) image shows the fabricated 3C-SiC microdisks with different radii sitting on silicon pedestals. The pedestal of the smallest microdisk is critically controlled to optimize the mechanical quality of the radial-breathing mode. The smallest microdisk is darker due to the carbon deposition in the SEM process. (b) and (c) illustrate the mode profiles for a whispering-gallery optical mode and the fundamental radial-stretching mechanical mode, respectively, with the insets showing the cross-section view. Both mode profiles are simulated by finite-element methods.}
\label{SEM}
\end{figure}

The devices we employed are cubic-type (3C) silicon carbide (SiC) microresonators sitting on silicon pedestals. The device fabrication process is described in \emph{Methods}. Figure \ref{SEM}(a) shows the fabricated devices of different radii with smooth sidewalls and fine-controlled undercuts. The fabrication process is optimized to produce smooth sidewalls, which are critical for minimizing the scattering loss of the optical modes. The device undercuts are optimized to reduce the clamping loss, which improves the mechanical qualities of the radial-breathing modes.

The microresonator exhibits whispering-gallery optical modes (Fig.~\ref{SEM}b) that produce radiation pressure along the radial direction to actuate the fundamental radial-breathing mechanical modes (Fig.~\ref{SEM}c), which in turn changes the cavity length and thus shifts the optical resonance frequency. The resulting dynamic backaction between the optical field and mechanical motion can be used to excite and probe the coherent mechanical motion, with efficiency dependent on the optomechanical coupling strength. For a microdisk optomechanical resonator with a radius of $R$, the optomechanical coupling coefficient scales as $\rm g_{om} \propto - {\omega_0}/R$, where $\omega_0$ represents the optical resonance frequency. The detailed simulations by the finite-element method (FEM) show that a SiC microdisk with a radius of 2~${\rm \mu m}$ and a thickness of 700~nm exhibits optomechanical coupling coefficients of ${\rm |g_{om}|/(2\pi)=89}$ and ${\rm 73~GHz/nm}$, respectively, for the fundamental and second-order transverse-electric-like (TE-like) modes, which correspond to a strong radiation pressure force of ${\rm |\hbar g_{om}|=59}$ and 48~fN produced by each photon, respectively. The FEM simulation indicates that the fundamental radial-breathing mechanical mode of the device exhibits an effective motional mass of ${\rm m_{eff} = 22}$~picograms. As a result, the vacuum optomechanical coupling rate, ${\rm g_0=g_{om} \sqrt{\hbar/(2m_{eff}\Omega_m})}$, is as large as ${\rm |g_0|/(2\pi)=42~kHz}$ for the fundamental TE-like modes in the device.

\subsection*{Optical Q characterization}

\begin{figure}[h]
\centering
\setlength{\belowcaptionskip}{-7pt}
\includegraphics[width=1\columnwidth]{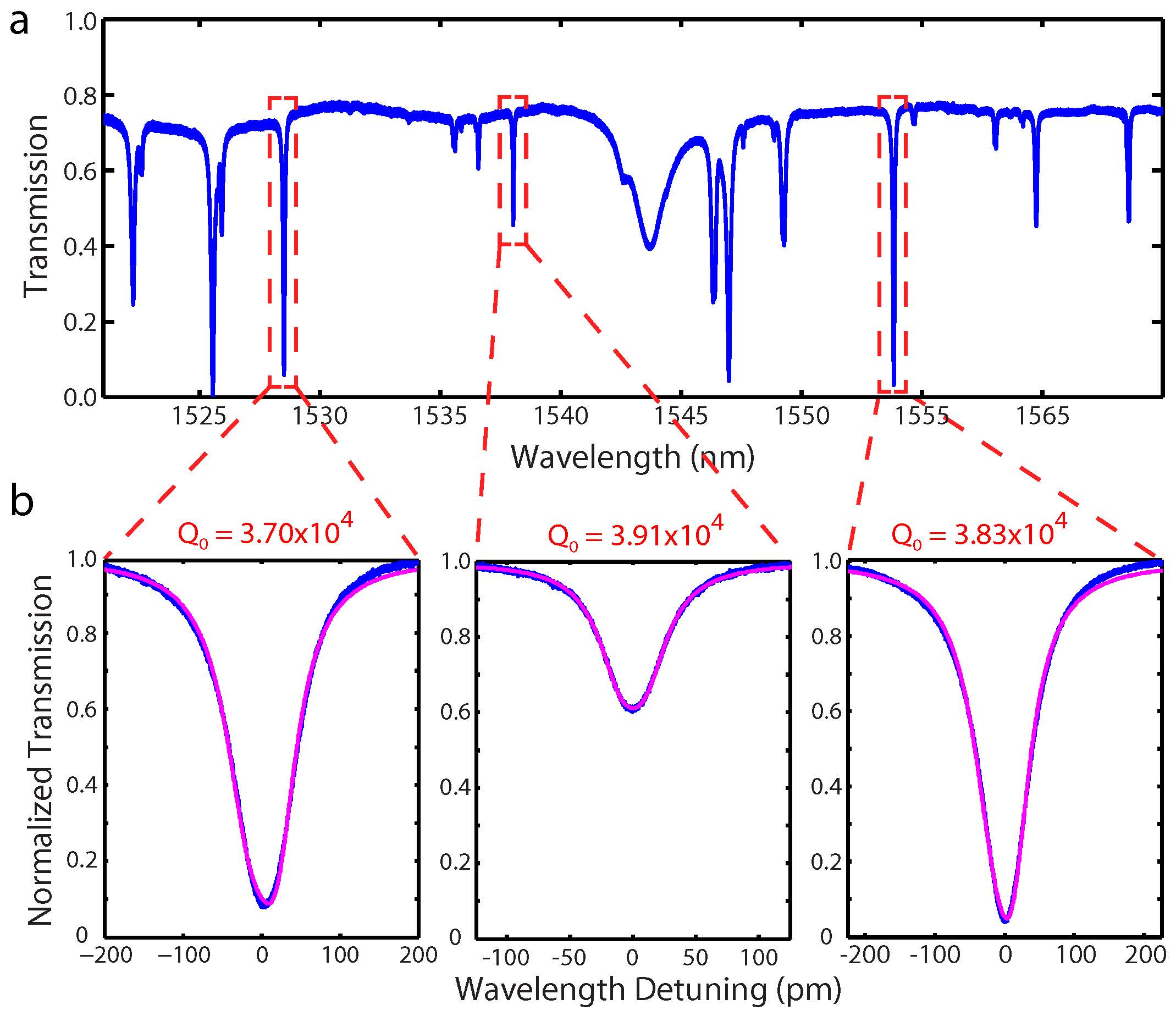}
\caption{ (a) Cavity transmission of a typical SiC optomechanical microresonator. (b) Three cavity modes have intrinsic optical qualities around $3.8 \times 10^{4}$, with experimental data in blue and theoretical fitting in red.}
\label{Transmission}
\end{figure}

The optical properties of devices are tested by a fiber-device coupling setup shown in Fig.~\ref{Setup}. A tunable laser is launched into the devices by evanescent coupling through a tapered optical fiber. The cavity transmission is coupled out by the same tapered fiber and then recorded by fast detectors. The laser wavelength is calibrated by a Mach-Zehnder interferometer. A typical cavity transmission trace is shown in Fig.~\ref{Transmission}(a) with multiple high-Q optical modes. Three optical modes from different mode families all show optical qualities around $3.8 \times 10^4$ (Fig.~\ref{Transmission}(b)). The coupling conditions of these modes can be easily tuned from under coupled, critical coupled to over coupled by tuning the fiber-device distance. For example, the cavity modes located around 1528~nm and 1553~nm are nearly critically coupled in this case.

\subsection*{Optomechanical excitation and sensing}

\begin{figure}[h]
\centering
\setlength{\belowcaptionskip}{-7pt}
\includegraphics[width=1\columnwidth]{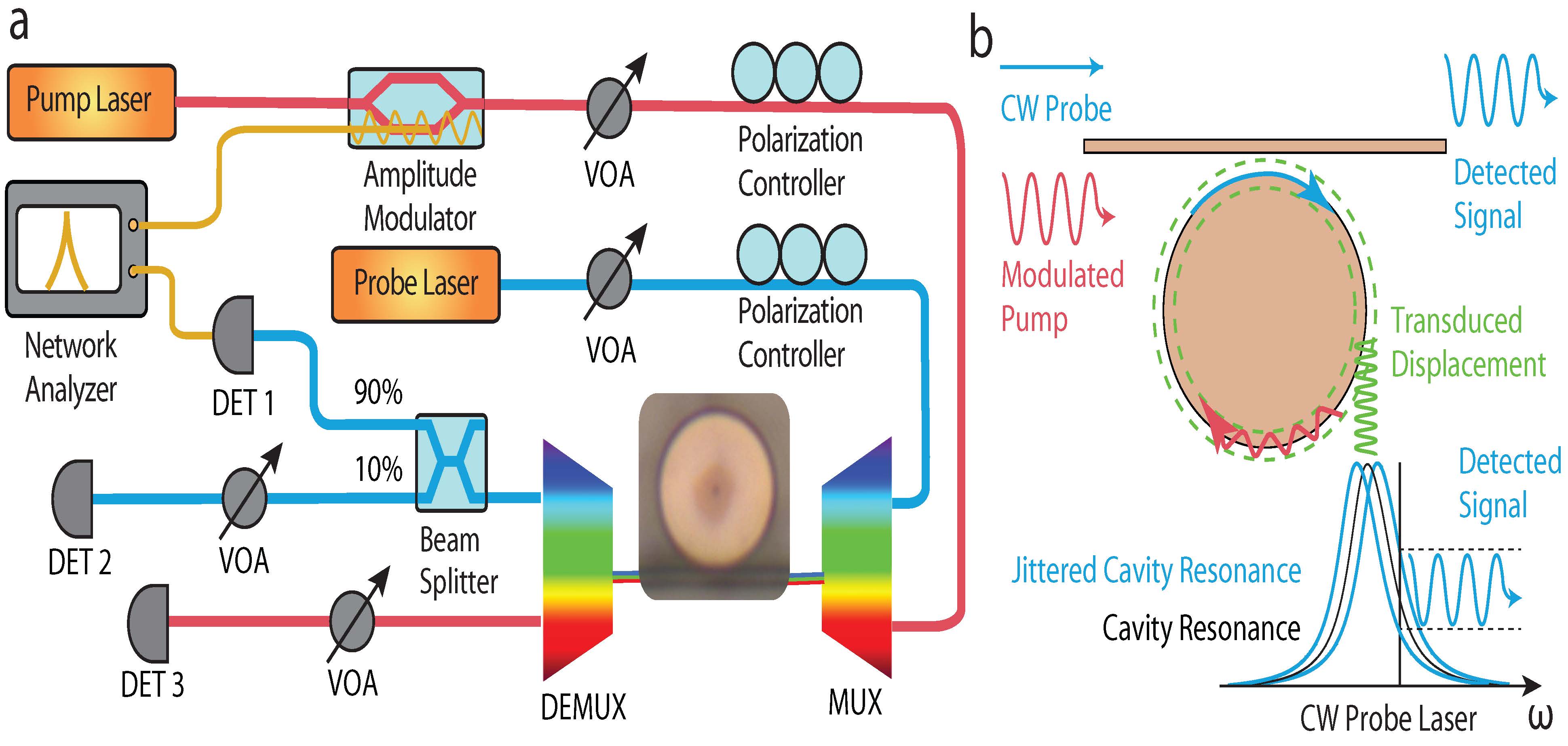}
\caption{(a) The experimental setup for the optical pump-probe scheme. VOA, MUX, and DEMUX represent variable optical attenuator, multiplexer, and demultiplexer, respectively. (b) An illustration of the pump-probe scheme.}
\label{Setup}
\end{figure}

The high optical quality of the whispering gallery modes, combined with the strong optomechanical coupling, enables efficient excitation and probing of the mechanical motion. To do so, we launch an optical wave (the pump wave) into a cavity resonance, with power sinusoidally modulated at a frequency around the mechanical resonance frequency. The operation principle is illustrated in Fig.~\ref{Setup}(b). A sinusoidal modulation of the optical power leads to a sinusoidally time varying radiation pressure that actuates the radial-breathing mechanical motion coherently via the strong optomechanical coupling. To probe such optomechanical excitation, we launch a weak continuous-wave optical wave (the probe wave) at a different cavity resonance. The coherent optomechanical excitation modulates the probe field inside the cavity via the optomechanical coupling, which is in turn transduced to the cavity output. Figure \ref{Setup}(a) shows schematically the experiment testing setup, with more detailed information given in the \emph{Methods}. The devices are tested at room temperature in the atmospheric environment.

A detailed analysis of the optomechanical dynamics shows that the modulated probe power, $\delta P_s(\Omega)$, at the modulation frequency $\Omega$, detected at the cavity transmission is given by
\begin{eqnarray}
 {\rm \frac{\delta P_s(\Omega)}{P_{\rm 0s}} = \left[ \frac{g_{\rm om}^2}{m_{\rm eff} \omega_0 \mathcal{L}(\Omega)} + 2\gamma_s \right] \delta U_p (\Omega) H_s(\Delta_s)}, \label{Eq_probe1}
\end{eqnarray}
where ${\rm \delta U_{p}(\Omega)}$ represents the modulated intra-cavity pump energy. ${\rm H_s(\Delta_s)}$ is the cavity transduction function of the probe mode. The detailed expressions of ${\rm \delta U_{p}(\Omega)}$ and ${\rm H_s(\Delta_s)}$ can be found in Ref.~\cite{Lin14}. Eq.~(\ref{Eq_probe1}) includes both optomechanical effect and optical Kerr effect. The first term describes the optomechanical response, with ${\rm \mathcal{L}(\Omega) \equiv \Omega_m^2-\Omega^2-i \Gamma_m \Omega}$ where ${\rm \Omega_m}$ and ${\rm \Gamma_m}$ are the frequency and damping rate of the mechanical mode, respectively. The second term containing ${\rm \gamma_s}$ describes Kerr nonlinear response, with ${\rm \gamma_s = c\omega_{\rm 0s} n_2 / (n_0^2 V_{\rm eff})}$ where ${\rm n_0}$ and ${\rm n_2}$ are the refractive index and Kerr nonlinear coefficient of SiC, respectively. ${\rm \omega_{\rm 0s}}$ is the resonance frequency of the probe mode and ${\rm V_{\rm eff}}$ represents the effective volume of the optical mode.

Our devices fall into the sideband unresolved regime, where the mechanical frequency is much smaller than the optical linewidth \cite{Aspelmeyer14}. In this regime, Eq.~(\ref{Eq_probe1}) can be simplified to
\begin{eqnarray}
 {\rm \frac{\delta P_s(\Omega)}{P_{\rm 0s}} = \left[ \frac{g_{\rm om}^2}{m_{\rm eff} \omega_0 \mathcal{L}(\Omega)} + 2\gamma_s\right] \frac{\delta P_d (\Omega)}{\Gamma_{\rm 0p}} \frac{2 \Gamma_{\rm es} \Gamma_{\rm 0s} \Delta_s}{\left[ \Delta_s^2+(\Gamma_{\rm ts}/2)^2 \right]^2}},
  \label{Eq_probe2}
\end{eqnarray}
where ${\rm \delta P_d (\Omega)}$ stands for the modulated pump power dropped inside the cavity. ${\rm \Gamma_{\rm 0p}}$ is the intrinsic photon decay rate of the pump mode. ${\rm \Gamma_{\rm 0s}}$ and ${\rm \Gamma_{\rm ts}}$ represent intrinsic and total photon decay rate of the probe mode, respectively. ${\rm \Gamma_{\rm es}}$ represents its external coupling rate. ${\rm \Delta_s = \omega_s - \omega_{0s}}$ is the laser-cavity detuning of the probe wave.

In the experiments, the optical mode is typically near critical-coupling conditions, ${\rm \Gamma_{\rm 0s} = \Gamma_{\rm es}}$, and the laser detuning for the probe mode is set around the half of total cavity linewidth ${\rm \Delta_s \sim \Gamma_{\rm ts}/2}$. As a result, Eq.~\ref{Eq_probe2} reduces considerably to
\begin{eqnarray}
 {\rm \frac{\delta P_s(\Omega)}{P_{\rm 0s}} = \left[ \frac{g_{\rm om}^2}{m_{\rm eff} \omega_0 \mathcal{L}(\Omega)} + 2 \gamma_s\right] \frac{\delta P_d (\Omega)}{2 \Gamma_{\rm 0p} \Gamma_{\rm 0s}}}.
  \label{Eq_probe3}
\end{eqnarray}
Equation (\ref{Eq_probe3}) clearly shows the linear dependence of the transduced probe signal on the optical qualities of the pump and probe modes. Moveover, it depends quadratically on the optomechanical coupling coefficient ${\rm g_{\rm om}}$ since the optomechanical effect not only drives the mechanical mode by the modulated pump beam, but also transduces the mechanical motion to the probe beam. Consequently, significant optomechanical coupling and high optical quality in the devices would lead to efficient optomechanical excitation and transduction by the pump and probe waves.

\begin{figure}[h]
\centering
\setlength{\belowcaptionskip}{-7pt}
\includegraphics[width=1\columnwidth]{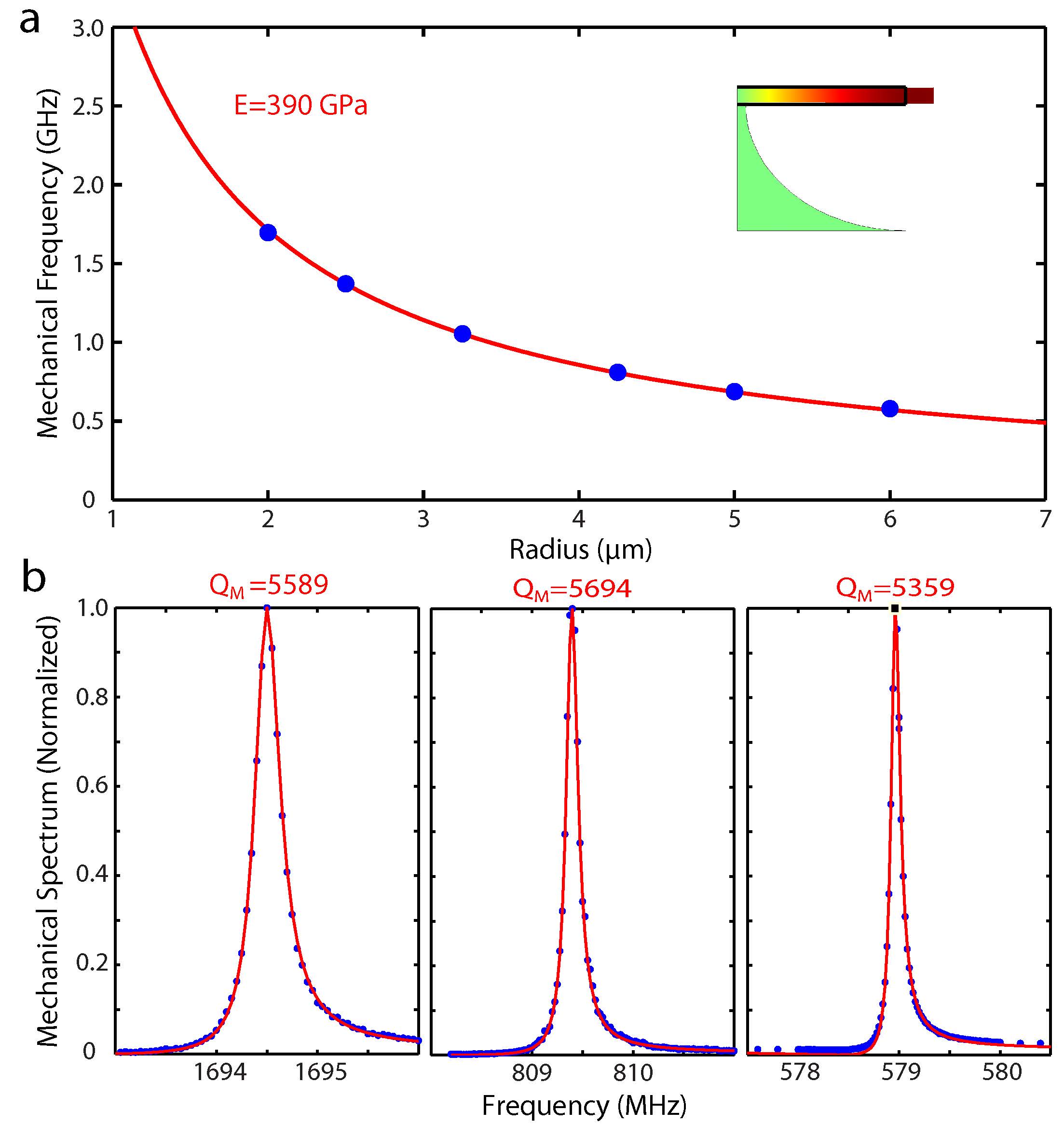}
\caption{(a) Mechanical frequencies of the fundamental mechanical radial-stretching modes are inversely proportional to the radii of the microdisks.  Experimental dots are in blue and the fitting curve is in red. Inset represents the displacement of a typical fundamental mechanical radial-stretching mode, with the geometrical edges outlined in black. (b) Normalized mechanical transduction spectra of the silicon carbide microdisks with radii being 2, 4.25, and 6 ${\rm \mu m}$, shown from left to right. Experimental dots are in blue and fitting curves are in red. The data are fitted by Eq.~\ref{Eq_probe2}. The silicon carbide microdisks maintain high mechanical Q factors around 5,500 for all the devices.}
\label{Mech}
\end{figure}

Equations (\ref{Eq_probe1})-(\ref{Eq_probe3}) show that, by scanning the modulation frequency, we can obtain the mechanical response of the radial-breathing mode. Figure \ref{Mech}(b) shows three examples of devices with different radii of 2, 4.25, and 6 ${\rm \mu m}$, respectively. The radial-breathing mechanical modes exhibit distinctive mechanical frequencies in these devices but all with a mechanical Q above 5000. The slight spectral asymmetry on the mechanical spectra is primarily due to the Fano-type interference between the narrow-band mechanical response and the broadband background of optical Kerr nonlinear response (see Eq.~(\ref{Eq_probe2})). A comparison of the recorded optomechanical spectra with the theory infers an optomechanical coupling coefficient of ${\rm |g_{\rm om}|/(2\pi) = (61 \pm 8)~{\rm GHz/nm}}$ for the 2 ${\rm \mu m}$ device. This value agrees closely with the theoretical expectation obtained from the FEM simulations. Moreover, we characterize the devices with different radii to map out the dependence of mechanical frequency. As shown in Fig.~\ref{Mech}(a), the mechanical frequency of the radial-breathing mode scales inversely with the device radius. Comparing the experimental data (blue dots) with the theoretical prediction (red curve), we infer the Young's modulus to be 390~GPa, which is consistent with previous measurements of 3C-SiC epitaxial films on silicon substrates \cite{Tong92}.

\begin{figure}[ht]
\centering
\setlength{\belowcaptionskip}{-7pt}
\includegraphics[width=1\columnwidth]{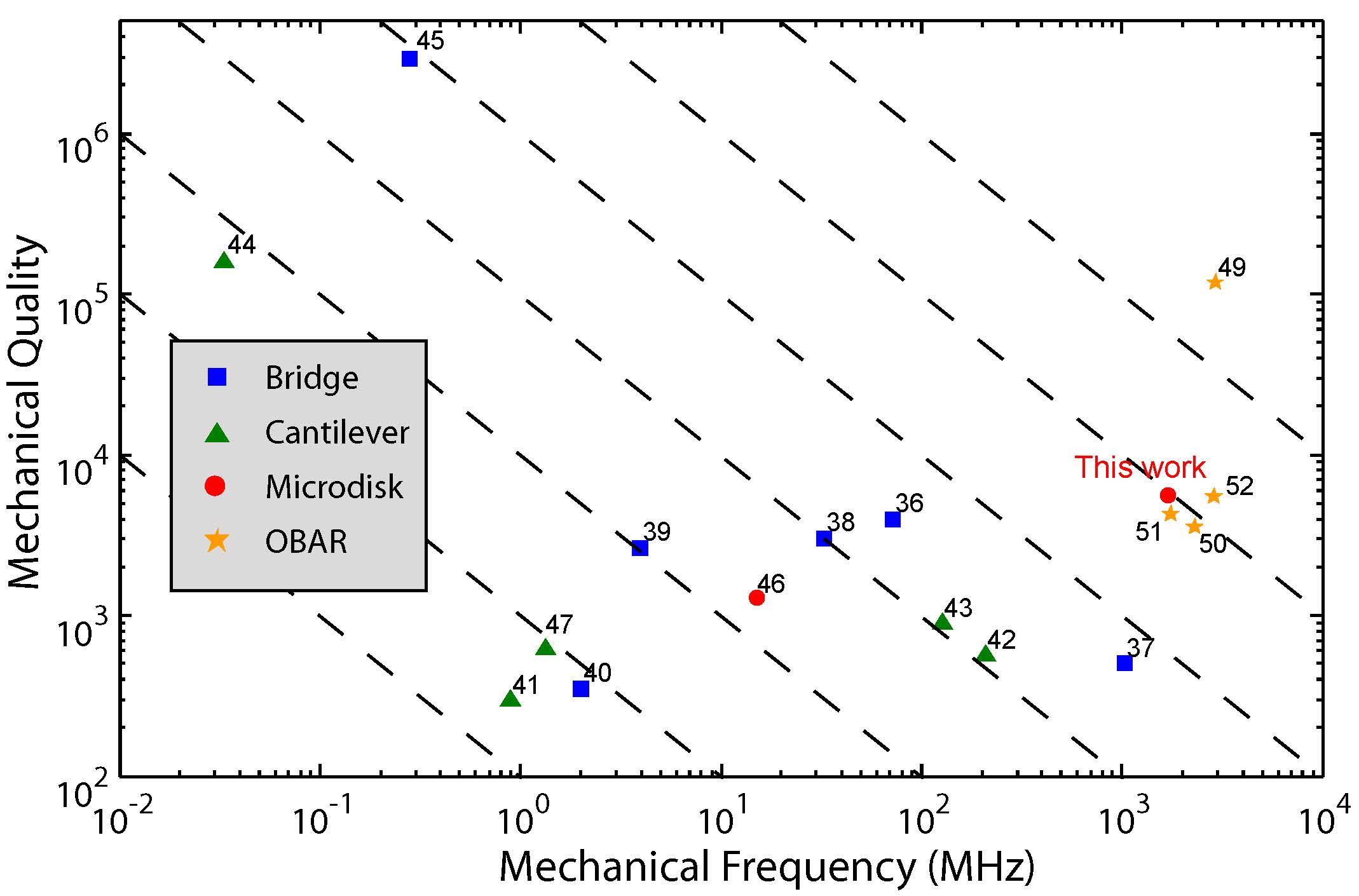}
\caption{The frequency-quality products of the SiC mechanical resonantors. Blue squares, green triangles, red circles, and yellow stars represent bridges, cantilevers, microdisks, and overtone bulk acoustic resonators (OBARs), respectively. The dashed black lines show the equal ${\rm f_m\cdot Q_m}$ product lines from $10^{14}$~Hz (top right) to $10^8$~Hz (bottom left).}
\label{fQ}
\end{figure}

One critical figure of merit for mechanical resonators is the ${\rm f_m \cdot Q_m}$ product, which quantifies the degree of decoupling of mechanical motion from the environmental thermal reservoir \cite{Aspelmeyer14}. Figure \ref{fQ} summarizes the ${\rm f_m \cdot Q_m}$ product reported to date for SiC micro/nanomechanical resonators  \cite{Ziaei11_1, Ziaei11_2, Lin12, Gong12, Yang01, Huang03, Ekinci04, Wiser05, Forster06, Jiang06, Azevedo07, Li07, Perisanu07, Yang14, Zhao14, Iacopi14}. In general, bridge- and cantilever-type SiC micro/nanomechanical resonators exhibit low ${\rm f_m \cdot Q_m}$ products, with a mechanical damping dominated by the mechanical clamping loss. To mitigate the clamping loss, high-order overtone-bulk-acoustic-resonator (OBAR) modes are employed to store mechanical energy over many mechanical wavelengths \cite{Ziaei11_1, Ziaei11_2, Lin12, Gong12}, which, however, requires a large device size significantly greater than the mechanical wavelength that seriously limits the device miniaturization and integration.

In contrast, our optomechanical resonators operate in the fundamental radial-breathing acoustic mode, with a small device size comparable to the mechanical wavelength. For example, the device with a radius of 2 ${\rm \mu m}$ exhibits a frequency of 1.69 GHz and a mechanical Q of 5589 (Fig.~\ref{Mech}(b)), which corresponds to a ${\rm f_m\cdot Q_m}$ product of ${\rm 9.47 \times 10^{12}}$ Hz. This product is among the largest values reported up to date of SiC devices \cite{Ziaei11_1, Ziaei11_2, Lin12, Gong12, Yang01, Huang03, Ekinci04, Wiser05, Forster06, Jiang06, Azevedo07, Li07, Perisanu07, Yang14, Zhao14, Iacopi14}, as shown in Fig.~\ref{fQ}. In fact, our device has the largest ${\rm f_m\cdot Q_m}$ product among whispering-gallery-type optomechanical microresonators made from various materials \cite{Vahala05,Favero10,Tang12,Wu13,Barclay14,Jiang12,Srinivasan13}. This value is still about an order of magnitude lower than the theoretical ${\rm f_m\cdot Q_m}$ product \cite{Kenny08, Ayazi11, Kenny13}, implying that the current limitation is not on intrinsic mechanical loss of SiC material, but on practical factors such as device etching, pillar clamping, and air damping. We thus expect improvement of the ${\rm f_m\cdot Q_m}$ product in the future after further optimization of the device structure and fabrication process.

\section*{Discussions}
We have demonstrated the first SiC optomechanical resonators in 3C-SiC microdisks that exhibit strong optomechanical coupling and excellent mechanical qualities, with a ${\rm f_m \cdot Q_m}$ product as high as ${\rm 9.47 \times 10^{12}~Hz}$. The high performance of the demonstrated devices infers that the SiC optomechanical devices are of great potential for metrology and sensing applications, particularly in detecting displacement, force, mass, and acceleration/rotation with high sensitivity. In combination with SiC's superior thermal property, chemical inertness, hand high breakdown voltage, SiC optomechanical devices are of great promise for applications in various harsh environments, such as those with high temperature, reactive chemicals, biological fluid, or high electric field \cite{Harris95, Azevedo07, Cimalla07, Zhuang05, Wright07, Oliveros13}, that are challenging for other device platforms.

On the other hand, the SiC optomechanical microresonators exhibit a mechanical frequency scalable by the device radius. In particular, the SiC microdisk with a radius of 2.5~${\rm \mu m}$ exhibits a mechanical frequency of 1.33~GHz (see Fig.~\ref{Mech}), which matches the zero-field splitting of spin ground states of the point defects in 3C-SiC \cite{Koehl11, Awschalom13}. Therefore, the high-Q collective mechanical mode is potentially able to coherently interact with the ground states of the defect spin via stress-induced coupling. This mechanism, in combination with the photon-spin coupling in SiC \cite{Awschalom14, Lu14} and photon-photon interaction via SiC's significant ${\rm \chi^{(2)}}$ and ${\rm \chi^{(3)}}$ nonlinearities \cite{Noda14, Lin14}, is of great potential to form a hybrid micro-/nano-photonic circuit that mutually couples photon, defect spin, and acoustic phonon for nonlinear optical, quantum optical, and optomechanical functionalities.

\section*{Methods}
\subsection*{Device fabrication}
The device structure we employed is cubic-polytype silicon carbide (3C-SiC) microdisks sitting on silicon pedestals. A high-definition electron-beam resist (ZEP520A) is used to pattern Chromium (Cr) mask with chlorine-based plasma by reactive-ion etching (RIE). The Cr mask is later used as a hard mask to etch SiC with fluorine-based plasma by inductively coupled-plasma RIE. The residue of Cr is then released by CR-14, a Cr etchant, and the silicon substrate is undercut by potassium hydroxide. The device is annealed afterwards at $\rm 1100~^{\circ}{C}$ for 2 hours. Figure \ref{SEM} shows the fabricated devices of different radii with smooth sidewalls and fine-controlled undercuts. More fabrication details can be found in Ref.~\cite{Lu14}.

\subsection*{Pump-probe setup}
The experimental setup is shown in detail in Fig.~\ref{Setup}(a). An intensive laser wave is sinusoidally modulated in amplitudes by a lithium niobate modulator. The frequency of modulation is scanned by a network analyzer. The pump laser is attenuated by a variable optical attenuator (VOA) to $\rm \sim 80~\mu W$. The probe laser is kept 10dB smaller than the pump beam by another VOA. The thermal effect is negligible for the operating powers in the devices. The polarization controllers are used to change the polarizations of the laser beams to the employed cavity modes. A coarse-wavelength-division-multiplexing (CWDM) multiplexer is used to combine the pump and probe beams and launch them into the cavity. The modulated pump beam drives the mechanical mode, with the mechanical displacement transduced to the jittering of the cavity resonance frequencies. The pump and probe beam are then separated by the CWDM demultiplexer. Detector 1, with 90\% transmission of probe beam, is collected by the network analyzer. The network analyzer scans the modulation frequencies and detects the signal at the same frequencies simultaneously. Detectors 2 and 3 are used for locking laser cavities to probe and pump modes, respectively. The optical modes we employed in the experiments are high order modes, which can be easily critically coupled by the current tapered fiber. The optomechanical coupling can be improved by accessing the fundamental modes through thinner tapered fiber or waveguide coupling.

\section*{Acknowledgements}
The authors thank Philip X.-L. Feng for helpful discussions. This work was supported by National Science Foundation under grant ECCS-1408517. It was performed in part at the Cornell NanoScale Science \& Technology Facility (CNF), a member of the National Nanotechnology Infrastructure Network.

\section*{Author contributions statement}
X.~L. and J.Y.~L. fabricated the devices and conducted the experiments. X.~L. analyzed the data. Q.~L. planned and supervised the project. All authors participated in the discussion of the results and the writing of the manuscript.

\section*{Additional information}
Competing financial interests: the authors declare no competing financial interests.

\newpage

\newpage

\newpage

\newpage

\newpage

\end{document}